\documentclass[letterpaper]{emulateapj}

\newcommand{\etal}{{\it et al.}}
\newcommand {\apgt} {\ {\raise-.5ex\hbox{$\buildrel>\over\sim$}}\ }
\newcommand {\aplt} {\ {\raise-.5ex\hbox{$\buildrel<\over\sim$}}\ }

\shorttitle{Weak Lensing Constraints on Q0957+561}
\shortauthors{Nakajima \etal}

\begin{document}

\title{Improved Constraints on the Gravitational Lens Q0957+561.\\
I. Weak Lensing}

\author{R. Nakajima}
\affil{Department of Physics and Astronomy, University of Pennsylvania,
Philadelphia, PA 19104 \\
Space Sciences Laboratory, University of California, Berkeley, CA 94720}

\author{G.~M. Bernstein}
\affil{Department of Physics and Astronomy, University of Pennsylvania,
Philadelphia, PA 19104}

\author{R. Fadely and C.~R. Keeton}
\affil{Department of Physics and Astronomy, Rutgers University,
Piscataway, NJ 08854}

\and

\author{T. Schrabback}
\affil{Argelander-Institut f\"{u}r Astronomie, Universit\"{a}t Bonn,
D-53121 Bonn, Germany\\
Leiden Observatory, Universiteit Leiden, NL-2333 CA Leiden, 
The Netherlands  
}

\begin{abstract}
Attempts to constrain the Hubble constant using the strong gravitational
lens system Q0957+561 are limited by systematic uncertainties in the 
mass model, since the time delay is known very precisely.
One important systematic effect is the mass sheet degeneracy, which arises
because strong lens modeling cannot constrain the presence or absence of a 
uniform mass sheet $\kappa$, which rescales $H_0$ by the factor
$(1-\kappa)$.  
In this paper we present new constraints on the mass sheet derived from 
a weak-lensing analysis of the Hubble Space Telescope imaging of a 
6 arcmin square region surrounding the lensed quasar.
The average mass sheet within a circular aperture (the strong lens 
model region) is constrained by integrating the tangential weak 
gravitational shear over the surrounding area.
We find the average convergence within a $30''$ radius around the lens
galaxy to be
$\kappa(<30'') = 0.166\pm0.056$ ($1\sigma$ confidence level), 
normalized to the quasar redshift. 
This includes contributions from both the lens galaxy and 
the surrounding cluster.
We also constrain a few other low-order terms in the lens potential
by applying a multipole aperture mass formalism to the 
gravitational shear in an annulus around the strong lensing region.  
Implications for strong lens models and the Hubble constant 
are discussed in an accompanying paper.
\end{abstract}

\section{Introduction}

Gravitational lensing allows for a measurement of the Hubble
constant $H_0$ that is independent of the ``distance ladder''
and is not susceptible to the peculiar velocities of the local universe
\citep{refsdal64}.
This unique opportunity is available for special lenses strong enough to
generate multiple images from a single source, and when this source has
intrinsic variability, such that a differential time delay can be
observed between the multiple images. 

The doubly-imaged quasar Q0957+561 is the first confirmed example of 
strong gravitational lensing [SL] \citep{Walsh79}.  The quasar is variable; 
with the time delay accurately determined to $<1\%$
between the two images \citep{kundic97, colley03, shalyapin08}, 
the lack of precision in obtaining $H_0$ from this system in past measurements
($\pm35\%$, 2$\sigma$) lies in determining the mass distribution of the lens 
\citep{BF99, keeton00}.

One well-known problem in uniquely determining the mass distribution 
(and hence $H_0$) is the ``mass-sheet degeneracy'' 
\citep{falco85, gorenstein88}.
The lens mass is modeled given the multiple images as constraints;
the lensed light is traced back from the multiple image according to 
a given lens model, and checked for a consistent, single source.
Unfortunately, the true angular position of the source is 
not observable and hence not constrained; the mass-sheet degeneracy 
occurs because the angular position of the source object is 
degenerate with the constant surface mass density, or the 
``mass sheet'' \citep{BF99}.  Thus, the 
mass sheet manifests itself as a uniform magnification of the 
(unobservable) source plane.

This degeneracy in the SL model can be lifted if the average
mass overdensity in the SL model region is known.  
This is done by using the aperture-mass weak gravitational lensing [WL] 
technique \citep{KS93}.
WL generates a subtle but coherent distortion of the background galaxies,
which can be used to infer the mass distribution.  
In particular, the integrated tangential shear around a circular aperture
non-parametrically determines the two-dimensional average mass overdensity
within the aperture.

The WL measurement suffers from its own mass-sheet degeneracy, because
the WL signal must formally be integrated over the 
surrounding region out to infinity to obtain the mass distribution.
While this is not practical, a realistic scheme to 
estimate the average mass sheet in a SL region is to 
integrate the WL distortions over an annular region surrounding the SL model. 
WL then measures the differential mass sheet $\Delta\bar\kappa$ of the
disk region defined by the inner and outer circle of the annulus, where
the average convergence within the large circle is small, and 
can be estimated from an assumed mass profile.  
The mass sheet for Q0957+561 has previously been measured in 
this way \citep{FischerProFit}.  

There are other sources of degeneracy in obtaining $H_0$ from SL.
The Q0957+561 lens is a galaxy which sits within a modest cluster, so 
the lens mass model must include both sharply and smoothly varying terms, 
corresponding to that of the galaxy and cluster, respectively.  
The cluster potential can be expanded in Taylor series; the second-order 
terms (the lowest order contributing to gravitational lensing) corresponds
to the mass sheet and constant shear within the SL region.
Past studies of the lens model have shown that the uncertainty in 
$H_0$ measured from
Q0957+561 is dominated by the degeneracy between the galaxy 
ellipticity and cluster shear \citep[and references therein]{keeton00}.
For many models, the parametric fits tend to converge to 
a large cluster shear ($\gamma_c\sim$0.1--0.3),
while the quasar host galaxy lensed images seem to imply 
a small shear ($\gamma_c\leq0.1$) 
\citep{keeton00}.
Additionally, \citet{kochanek91} and subsequent papers have shown that
the third-order terms cannot be neglected for the Q0957+561 cluster.

The goal of this paper is to constrain the smoothly varying components
of the lens, i.e., the mass-sheet and a few of the lowest-order 
cluster terms, using WL analysis.  The convergence and shear pattern 
multipoles relative to an aperture are related to each other 
in a simple manner \citep{SB97, BN08}, and this relation has been 
applied to the newly acquired images from Hubble Space Telescope (HST) 
Advanced Camera for Surveys (ACS).
The combined lens model from strong lensing analysis
of the system is discussed in an accompanying paper \citep{fadely09},
as well as its implication for $H_0$.

\section{Theoretical Background}

\subsection{Time Delay and Hubble Constant}

When the gravitational field of a lens object is strong enough,
the light path can be sufficiently perturbed such that it 
generates multiple images of a single light source.
The bending of two or more separate light paths arriving at the observer
is determined from the lens configuration; i.e., 
the distances between observer, lens and source, as well as
the mass distribution of the lens.
The SL observations consist of (dimensionless) angular separations, redshifts, 
and relative magnitudes of the multiple images, 
and have no distance or mass scales directly associated to them.
A measurement of the time delays between
the multiple images provide an absolute time/distance scale to 
the lensing system \citep{refsdal64}.

Two light paths originating from the same object 
undergo different time delays due to differences in
(a) their overall path lengths, and
(b) the depth of gravitational potential as they cross the lens.
The time delay (with respect to a straight path from {\boldmath$\beta$}, 
the true, unobserved angular position of the source, with no intervening lens) 
for a single light path traveling through {\boldmath$\theta$} in the lens plane is
\begin{equation}
  t(\mbox{\boldmath$\theta$})
  = \frac{(1+z_{\rm lens})}{c}\,\frac{D_{OL}D_{OS}}{D_{LS}}\,
  \left[\frac{1}{2}(\mbox{\boldmath$\theta$}-\mbox{\boldmath$\beta$})^2-\psi(\mbox{\boldmath$\theta$})\right]
\end{equation}
where the delay due to the path difference is proportional to 
$(\mbox{\boldmath$\theta$}-\mbox{\boldmath$\beta$})^2/2$, and 
the other term represents the Shapiro time delay, where $\psi$ is 
the dimensionless, projected and scaled gravitational potential of the lens:
\begin{equation}
  \psi(\mbox{\boldmath$\theta$}) \equiv 
  \frac{D_{LS}}{D_{OL}D_{OS}}\frac{2}{c^2}
  \int\Phi(D_{OL}\mbox{\boldmath$\theta$},\ell)d\ell.
\end{equation}
Here, $\Phi$ is the ordinary 3D Newtonian potential,
$\ell$ is the distance along the line-of-sight, and 
$D$ indicates angular diameter distances between observer ($O$),
lens ($L$) and source ($S$).
The absolute delay $t(\mbox{\boldmath$\theta$})$ is not observable, but the 
difference $\Delta t = t(\mbox{\boldmath$\theta$}_1) - t(\mbox{\boldmath$\theta$}_2)$ is,
where $c\Delta t$ is proportional to $H_0^{-1}$ from the definition of
angular diameter distances.
Hence, we see that 
the Hubble constant $H_0$ can be obtained from the geometry of the 
lens system if $\Delta t$ is measured and the lensing potential 
$\psi(\mbox{\boldmath$\theta$})$ is known.  
In general, $\psi$ is obtained from fits to parameterized lens 
models \citep[\protect{\it e.g.,}][]{BF99,keeton00},
where the best fit to the strongly lensed images is chosen.

\subsection{Mass Sheet Degeneracy}
\label{sec:msdegeneracy}

A fit to a parameterized lens model $\psi$ does not provide
us with a unique solution, however.  
A uniform mass sheet cannot be constrained by (and hence not included in)
the SL models.
Specifically, for a lens potential $\psi(\mbox{\boldmath$\theta$})$ and 
source position $\mbox{\boldmath$\beta$}$ that satisfies the SL constraints, 
\begin{eqnarray}
  \psi'(\mbox{\boldmath$\theta$}) 
  &=& \frac{1}{2}\kappa_0|\mbox{\boldmath$\theta$}|^2 + (1-\kappa_0)\psi(\mbox{\boldmath$\theta$}) \\
  \mbox{\boldmath$\beta$}'
  &=& (1-\kappa_0)\mbox{\boldmath$\beta$}
\end{eqnarray}
also satisfies the constraints, where $\kappa_0$ is 
the constant convergence (see Eq.(\ref{eq:convergence})) from 
the additional uniform mass sheet.
Under this degenerate transformation, the time delay is
\begin{equation}
  (\Delta t)' = (1-\kappa_0) \Delta t,
\end{equation}
requiring the Hubble parameter to rescale by $H_0' = (1-\kappa_0)H_0$.

The lensing convergence $\kappa(\mbox{\boldmath$\theta$})$ is a normalized 
surface density, and is defined as
\begin{equation}
  \nabla^2\psi(\mbox{\boldmath$\theta$})
  = 2\kappa(\mbox{\boldmath$\theta$})=2\frac{\Sigma(\mbox{\boldmath$\theta$})}{\Sigma_{\rm crit}}
\label{eq:convergence}
\end{equation}
where $\Sigma(\mbox{\boldmath$\theta$})$ is the surface mass density, and 
the critical surface density is determined from the lens and source redshifts,
and the assumed cosmology:
\begin{equation}
  \Sigma_{\rm crit}
  = \frac{c^2}{4\pi G} \frac{D_{OS}}{D_{OL}D_{LS}}.
\end{equation}
A SL region which can generate multiple images has
$\Sigma\apgt\Sigma_{\rm crit}$, or $\kappa\apgt 1$, while WL 
techniques are generally valid only in a region where $\kappa \ll 1$.

To remove the $(1-\kappa_0)$ degeneracy in $H_0$, one must obtain
the uniform mass sheet $\kappa_0$ not included in the SL model.
The overall average mass sheet ($\bar\kappa$) can be constrained via WL, 
which constrains the sum of $\bar\kappa_{\rm SL}$ and $\kappa_0$,
where $\bar\kappa_{\rm SL}$ is the average convergence of the SL model
lens mass distribution.  The WL information is obtained 
over a field wider than the area where $\psi(\mbox{\boldmath$\theta$})$ is modeled, 
and its procedure is described in \S\ref{sec:multipoles}.

\subsection{Cluster Model}
\label{sec:clustermodel}

The Q0957+561 lens system consists of the primary lens galaxy G1 at redshift 
$z_{\rm lens}=0.355$, the relatively weak cluster which contains G1, and 
the quasar itself which lies at $z_{\rm src} = 1.41$.
The lens model in the SL analysis consists of a mass concentration 
with near elliptical symmetry representing the galaxy, with 
additional cluster potential components that are Taylor expanded 
in position to the third order
\citep{BN08, fadely09}.
The expansion is defined within $r\leq R_{SL}$ centered at the 
central peak of G1, and 
is fully general to order $(r/R_{SL})^3$, where
$R_{SL}$ is the radius of the circular boundary where the SL modeling
takes place:
\begin{eqnarray}
\lefteqn{ \psi({\bf r}) = (1-\kappa_c) \times }\nonumber \\
&& \left[\psi_g({\bf r}) +{\rm Re}\left(
\frac{\gamma_c}{2}r^2 e^{-2i\phi} +
\frac{\sigma_c}{4}r^3 e^{-i\phi} +
\frac{\delta_c}{6}r^3 e^{-3i\phi} \right)\right] \nonumber\\
&& \ \ \ \ \ \ \ \ \ + \frac{\kappa_c}{2}r^2 
\ \ \ \ \ \ \ \ \ \ \ \ \ \ \ \ \ \ \ \ \ \ \ \ \ \ \ (r\leq R_{SL})
\label{eq:galcluster}
\end{eqnarray}
where ${\bf r}\equiv(x,y)\equiv(r,\phi)$ is the transverse distance in the 
lens plane 
(and is equivalent to $\mbox{\boldmath$\theta$}$ in the previous sections),
$\psi_g({\bf r})$ is the galaxy potential, 
and $\kappa_c$, $\gamma_c$, $\sigma_c$ and $\delta_c$ 
are the (complex) constants corresponding to the cluster mass sheet,
constant shear, internal mass dipole moment, and $m=3$ shear moment, 
respectively.
The complex notation corresponds to the two-dimensional
description of the gravitational potential such that, {\it e.g.}, 
the derivative is
$$
\partial \equiv \frac{\partial}{\partial x} 
             + i\frac{\partial}{\partial y}
= e^{i\phi}\left(\frac{\partial}{\partial r} 
  + \frac{i}{r}\frac{\partial}{\partial \phi}\right).
$$
The potential and the mass sheet $\kappa_c$ are always real.
The first four terms in the square bracket are those modeled in SL,
and are multiplied by $(1-\kappa_c)$ 
due to the mass sheet degeneracy associated with the analysis;
$\kappa_c$ is the mass sheet that is not part of the SL model, and
cannot be constrained from SL model within $r<R_{SL}$ alone, due to
degeneracy (\S\ref{sec:msdegeneracy}). 
The four quantities $\kappa_c$, $\gamma_c$, $\sigma_c$ and $\delta_c$
are constrained by WL measurements from data in 
$R_{SL}<r<R_{\rm max}$ and the multipole formalism (\S\ref{sec:multipoles}).
The inner and outer annular radii 
were chosen to be $R_{SL}=30''$ and $R_{\rm max}=186''$ for our analysis, 
such that we consider a region sufficiently removed from the SL region, 
but with reasonable WL signal from the cluster.

\section{Observations and Data}

\subsection{HST/ACS Imaging}

\begin{figure}
\plotone{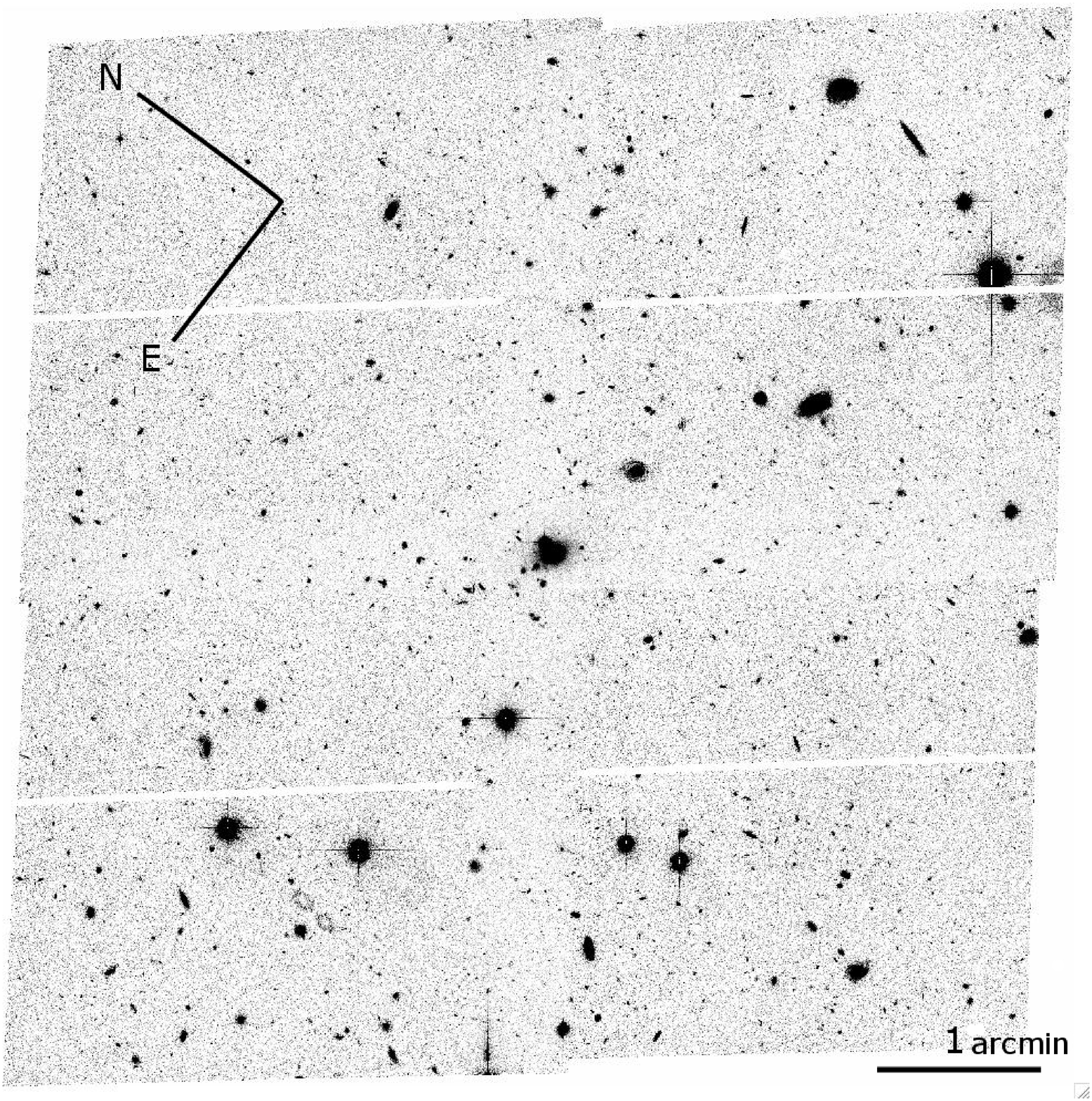}
\caption[The $6'\times6'$ combined image of the 0957+561 field in ACS F606W.]
{The $6'\times6'$ combined image of the 0957+561 field in ACS F606W.
Each quadrant is a single pointing (with pixel scale dithering) which 
has been imaged to a depth of 7.5ks, while all four quadrants overlap
in the central $30''$ region to yield a depth of 30ks.
The chip gap is apparent in each of the quadrants.
The orientation and the image scale are as labeled.
The apparent shear in the FOV is from correcting for the image distortion
due to the ACS focal plane located off-center from the telescope axis.
\label{fig:q0957field}}
\end{figure}

The images for the cluster around the double quasar 
Q0957+561 were taken as a $2\times 2$ mosaic of HST ACS 
pointings of the Wide Field Channel (WFC) detector on October 10--13, 2005
(Fig.~\ref{fig:q0957field}).  
Each ACS WFC pointing has $202''\times202''$ field of view (FOV), 
resulting in a $6'\times6'$ mosaic image.
Each pointing has 4 orbits (4 $\times$ 1920s) in the F606W 
(Broad V) filter and 2 orbits (2 $\times$ 1880s) in the F814W
(Broad I) filter; 
the mosaic has the central $30''$ region overlapped
such that the region around the quasar images have 
four times the depth (30ks and 15ks for F606W and F814W, respectively)
for studies of SL features \citep{fadely09}.

The HST WFC is a mosaic of two $4096\times2048$ pixel charge-coupled devices
(CCDs) of approximately $0.05''$/pixel.
Each orbit is split into four dithered exposures to sample the 
diffraction-limited point-spread function (PSF) and to remove 
defective pixels and cosmic rays in the combined image.

\subsection{Object Catalog}

\begin{figure}
\plotone{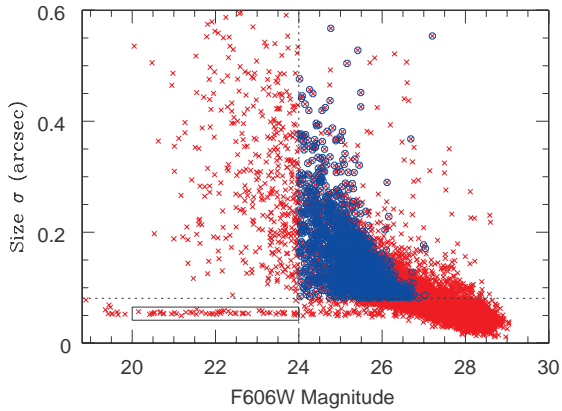}
\caption[Size-magnitude diagram of all detected objects from
native {\tt GLFit}.]
{Size-magnitude diagram of all detected objects from
native {\tt GLFit}.
The boxed region shows a portion of the stellar locus with
no saturation and no galaxy contamination.
The stellar size  $\sigma$ is approximately the size of the 
ACS WFC pixel, $0.054''$, where $\sigma$ is the characteristic 
width scale for the best-fit gaussian (for objects with elongated shapes,
$\sigma$ is the geometric mean of the two characteristic widths).
The crosses are all detected objects, while the
circles indicate objects which satisfy the size, magnitude, flags, and
signal to noise (S/N)$>25$ cut for use in WL analysis.  
The size and magnitude cuts are 
indicated by a dotted horizontal and vertical line, respectively.
\label{fig:size-mag}}
\end{figure}

We created an initial stacked mosaic image using {\tt MultiDrizzle}
\citep{multidrizzle}, where we refine relative shifts between 
individual exposures by cross-correlating the positions of compact sources 
in overlapping regions.  Objects were detected in this image using 
{\tt SExtractor} \citep{Bertin}.  
We took the initial {\tt SExtractor} detection parameters and 
refined the centroid, size and shape of each object using a native {\tt GLFit} 
\citep[\S\ref{sec:eglmethod};][]{NB07}.
The refined parameters determined here are still those which are
convolved with the PSF, therefore the size-magnitude 
diagram allows us to determine the stellar locus 
(Figure~\ref{fig:size-mag}).  
The stellar objects have an average full-width half maximum (FWHM) of 
$0.13\arcsec$ in F606W, or a corresponding $\sigma=0.054\arcsec$,
where $\sigma$ is the characteristic width of the best-fit gaussian.

Once the stars and galaxies were separated, 
the true (pre-PSF) galaxy shapes were obtained by performing
a deconvolution {\tt GLFit}, based on an interpolated PSF
(\S\ref{sec:psfinterp}).  
Both the native and deconvolution {\tt GLFit} are done as a 
``multifit''---an individual object is fitted simultaneously
over pixel data from each exposure rather than from the combined 
image---hence the measured shapes are not affected by distortion or 
aliasing induced by the image-combining process (\S\ref{sec:multifit}).

\section{Weak Lensing Analysis}

\subsection{EGL Method}
\label{sec:eglmethod}

For our WL analysis, we employ the Elliptical Gauss Laguerre (EGL) 
method of shape measurement and shear estimation \citep{BJ02, NB07},
implemented as {\tt GLFit}; we provide here a brief description.
In characterizing their shapes, star and galaxy images are decomposed into 
2d Gauss-Laguerre (GL) basis functions \citep{BJ02}.  In the EGL method,
the GL basis functions are described upon a ``basis ellipse,'' i.e., 
a sheared and stretched coordinate system such that a unit circle
appears as an ellipse of chosen size, shape or orientation; and 
so that the lowest order GL function is an elliptical 2d Gaussian.
The combination of the basis ellipse and coefficients to the GL functions 
then fully describes the image.

The choice of basis ellipse varies depending on its purpose.
In describing the PSF, we choose a common basis ellipse: a circle 
with radius of the average PSF size.
The PSF variation across the FOV can then be traced through the variation
in the GL coefficients, and allows for a simple description of the necessary 
interpolation.

In contrast, when measuring galaxy shapes, the basis ellipse is chosen
such that (in the simplified case of no PSF smearing) the lowest order 
GL function is the best-matched 2d gaussian ellipse to the galaxy image.
This basis ellipse then describes the shape (ellipticity) of the galaxy.
In the observed image, the true galaxy shape has been convolved with the PSF.
So our fitting does a simple deconvolution by modeling the true galaxy shape
with a GL expansion, and adjusting the coefficients so that convolution 
of this model with the PSF best matches the data \citep{NB07}.

\subsection{PSF Interpolation}
\label{sec:psfinterp}

In order to obtain the true galaxy shapes,
the PSF must be known at the position of the galaxy image.
This is done by selecting the stars in the image and 
interpolating the shape at the galaxy position.
There are two problems  that make this difficult for our images.
First, the PSF spatial variation is not stable over time:  
HST is known to ``breathe'' going in and out of the earth's shadow, 
causing its focus to vary on time scales of a single orbit ($\sim1.5$ hour).  
Hence it is unlikely that a single PSF spatial variation model
would be valid for all exposures \citep[and references therein]{Schrabback07}.
Second, there are not enough stars ($\sim$20/chip) in our cluster 
image to create a reliable model of the PSF variation across every exposure.

To circumvent the aforementioned difficulties, we use 
publicly available ACS WFC stellar field images to create the models of
the PSF variation for every exposure.  
A given PSF spatial pattern is highly reproducible, since the spatial 
variation of PSF patterns in different exposures is caused by the 
thermal breathing of the HST focus. 
If the few PSF available on a given exposure can constrain a portion 
of the pattern, the rest of the PSF pattern can be predicted
\citep{rhodes05, jee07}.

\begin{figure*}
\plottwo{plot_rank.epsi}{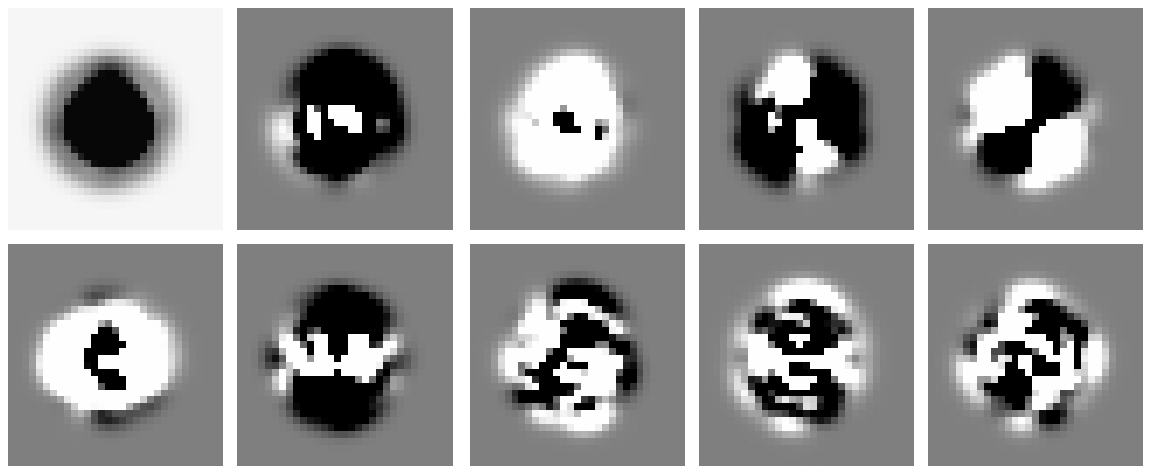}
\caption[(a) The variance of each of the principal components 
from each of the CCD chips.
(b) The first ten principal components of PSF in chip 1.]{
(a) The variance of each of the principal components from each of the 
CCD chips.  The dotted line indicates the last major gap in the principal
component variance.  
There are seven principal components to the left of this line.
(b) From top left to bottom right: 
the first ten principal components of PSF in chip 1.  
\label{fig:principalcomponents}}
\end{figure*}

The noise in the interpolation is minimized by utilizing 
principal component analysis (PCA).  Our procedure is as follows:
\begin{enumerate}
\item Obtain the stellar field (SF) images available from the HST archive.
There were 184 archival F606W exposures of dense stellar fields;
the criteria for selection are described in \citet{Schrabback07}.
\item The PSF anisotropy is measured from each star 
in terms of the anisotropy kernel $q$ \citep{Schrabback07}, 
and its variation across the field-of-view is characterized by a 
third-order polynomial fit.
\item We then choose, for each of our Q0957 exposure, the SF image that 
best matches the anisotropy pattern of the stars.
The ``best-fit'' stellar field is identified by that which yields the minimal 
$$
\chi^2_{\rm SF} 
= \sum_{i=1}^{N_{\rm stars}} [q_i - q_{\rm model}^{\rm SF}(x_i,y_i)]^2
$$
for a given exposure, where
$q_i$ is the kernel for the $i$th star in our exposure (which have 
very few stars), and $q^{\rm SF}_{\rm model}(x_i,y_i)$ 
is the interpolated kernel at the position of the $i$th star 
for the matching stellar field.
\item For the PSF interpolation, obtain the PSF GL coefficients by
(a) dividing each stellar field exposure into a $8\times8$ grid, and then
(b) fitting all stars within the grid to a constant PSF model for that cell,
resulting in 64 PSFs per exposure (32 PSFs per chip).
We choose to average the PSF in this way because the ACS WFC 
undersamples the PSF; by using multiple PSF images within the grid,
the PSF is effectively ``dithered'' to obtain better sampling for the fit.
We then have, for a given stellar field $f$ and within every grid cell $g$, 
a GL coefficient vector ${\bf b}^{fg} = \{b^{fg}_i\}$ describing the PSF,
where $i$ runs over the GL indices.  All PSF coefficients from $\{f,g\}$
are described over a common basis ellipse, which we take to be a circle 
with radius equal to the average size of the PSF.
\item Extract the principal components ${\bf b}^\alpha = \{b^\alpha_i\}$
of the collection of PSF GL coefficients $\{{\bf b}^{fg}\}$ \citep{JarvisPCA}.
Express the PSF in terms of the PCA vectors,
${\bf b}^{fg} = \sum_\alpha \beta^{fg}_\alpha{\bf b}^\alpha$,
where $\alpha$ runs over the PCA vectors.
The PSF is now described in terms of 
PCA coefficients $\mbox{\boldmath$\beta$} = \{\beta_\alpha\}$ 
instead of the GL coefficients ${\bf b} = \{b_i\}$.
\item Truncate the PCA coefficients by using only the 
major principal components which corresponds to the PSF (spatial and 
temporal) variation, but not to the noise.  
These components are identified by ranking the variance of each, and 
identifying a gap in the variance (see Fig.~\ref{fig:principalcomponents}).  
Most of the variation in PSF is described
in the first several terms in $\alpha$.
\item Obtain a third-order, two-dimensional polynomial fit 
across the image for the relevant PCA components $\alpha$
for each field $f$, such that the interpolated PCA coefficients
are described as
$$
\tilde\beta^{fg}_\alpha 
= \sum_{0\leq(n+m)\leq 3} c_{\alpha,nm}^{f} x_g^n y_g^m
$$
where 
$\chi^2=\sum_g(\beta^{fg}_\alpha - \tilde\beta^{fg}_\alpha)^2$ is minimized,
$c_{\alpha,nm}^f$ is the polynomial coefficient for $\alpha$ in field $f$,
and $(x_g,y_g)$ are the grid center coordinates.
Each PCA description of the PSF on every grid 
($\mbox{\boldmath$\beta$}^{fg}$) was visually inspected;
obvious failures in the PSF descriptions (due to lack of sufficient
stars within the grid) were eliminated before being used in this fit.
\item The PSF at any location $(x,y)$ over any field $f$ can then be
modeled using the known principal components and 
its 2-dimensional polynomial fit coefficients across the given exposure.
\end{enumerate}

Figure~\ref{fig:principalcomponents} shows the variance of 
each of the PSF principal components, for each WFC ACS CCD.
The zeroth component is the average PSF shape over all grid cells and 
exposures.  We find that the first seven components constitute the 
primary variance and are sufficient to describe the PSF variation; 
from visual inspection, the residual is consistent with $\sim1\%$ noise 
with respect to the peak.
These seven components are used to model the PSF across the exposure.

\subsection{Galaxy Multifit}
\label{sec:multifit}

Once the PSF is known at a given galaxy location, we perform the
deconvolution {\tt GLFit} to obtain the true galaxy shape.  
We utilize the multiple exposures taken of each galaxy without combining them.
The ``multifit'' technique performs simultaneous fitting over 
individual exposures, each with distinct PSFs, assuming a single 
``true'' galaxy model, and the fitting done with the convolved GL 
basis functions of each exposure to their respective PSFs.
The simultaneous fitting is iterated over the basis ellipses 
to obtain the best-fitting true galaxy shape.

The pixel image is distorted with respect to the true image, so
distortion correction is necessary within the multifit procedure.
The galaxy model, described in sky coordinates, is fitted to 
the pixel flux information via a pixel-to-sky coordinate map, which 
corrects for optical distortions in the pixel image.
The pixel map is based on a known, stable solution of 4th order polynomial
in combination with a supplementary look-up table 
\citep{anderson02, anderson06}, to which a linear correction (rotation,
translation, shear and uniform scale) is applied to fit to the 
USNO-B catalog \citep{usnob} for alignment and absolute astrometry.
Bad pixels, such as those affected by cosmic rays or saturation, are not
used in the fitting procedure.  The bad pixel masks for each exposure were 
generated using the {\tt multidrizzle} package \citep{multidrizzle}.

\subsection{Galaxy Selection for Weak Lensing}

\begin{figure}
\plotone{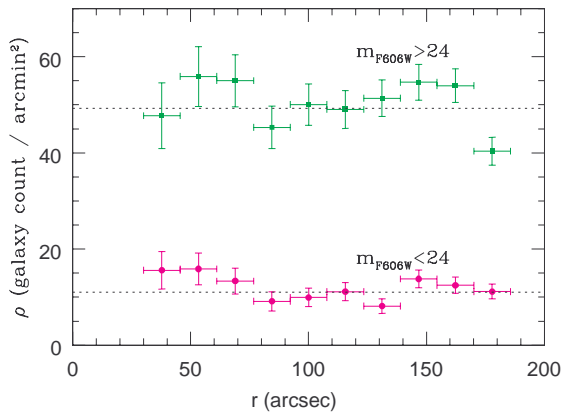}
\caption[Weighted galaxy number density as a function of radial distance 
from the central galaxy of the cluster.]
{Weighted galaxy number density as a function of radial distance 
from the central galaxy of the cluster.  
The weighted number density for use in shear estimating \citep{BJ02}
were essentially identical to the straight-forward number density.
The error bars are based on Poisson statistics.
The dimmer $m_{\rm F606W}>24$ objects are used for weak lensing analysis
(see text), where
both set of objects have the size, significance and flag cuts applied.
The dotted lines are the average weighted galaxy density for 
the region $75''<r<186''$, where the density averages to 
11.1 and 49.2 arcmin$^{-2}$ for the bright and dim galaxies, respectively.
The brighter objects ($m_{\rm F606W}<24$) have an excess of $\sim(30\pm15)\%$
in galaxy density for $r<75''$ from the cluster brightest galaxy G1;
the fainter objects (used for WL) appear to have an excess of 
$\sim(10\pm10)\%$.
\label{fig:rgalcounts}}
\end{figure}

Once the object catalog has been generated, we select the galaxies
for WL analysis by imposing the following cuts:
\begin{itemize}
\item Size.  The stellar contamination is removed by choosing objects
which have characteristic size $\sigma$ which are $\geq1.5$ times that 
of the PSF.
Our cut is conservative  to ensure PSF deconvolution
to be in the range where it performs well \citep{NB07}.
\item Significance (S/N).  From our performance analysis of {\tt GLFit}
\citep{NB07}, we know that the deconvolution starts to produce
biases below S/N of 20 to 40, depending on the galaxy size with respect
to the PSF.  Here we choose the cut to be S/N $>25$.
\item Magnitude.  Since we only want the background galaxies to the
cluster for the lensing analysis, foreground cluster member contamination 
must be minimal.  Lacking redshift information, and
unable to utilize color information for this purpose for reasons 
listed below, we cut the brightest objects
($m_{\rm F606W}<24$) as foreground.  By plotting the galaxy number
density as a function of the radius from the central brightest galaxy
of the cluster (Figure~\ref{fig:rgalcounts}), we estimate that there is 
at most 10\% cluster member contamination for $r<75''$ from 
the cluster center for the $m_{\rm F606W}>24$ objects.
\item Flags.  Any object whose {\tt GLFit} flags indicate that the shape
has not been measured (an unsuccessful fit to the basis ellipse) is 
rejected from the catalog.
\end{itemize}
After these cuts, 1866 galaxies remain for our WL analysis, or 
galaxy number density of $50$ arcmin$^{-2}$ (Fig.~\ref{fig:rgalcounts}).

We have not included a color cut (for a rough exclusion of cluster members)
for the following reason:  Although several of the red-sequence
cluster members of known redshifts have the predicted 
${\rm F606W} - {\rm F814W}$ color of 1.0 \citep{bc93} in our color-magnitude
plot, no peak was found in the galaxy count at this color. 
This is because (1) majority of the cluster members are blue,
and (2) the expected member galaxy count in this halo is small.  
The first point is verified from $>50\%$ of cluster members of known redshifts 
which are bluer than red-sequence, with ${\rm F606W} - {\rm F814W}$ 
colors in the range 0.3--0.8.  Many of these galaxies were found
to have spiral morphology \citep{angonin94}.  
The Q0957 cluster mass (and hence the number of its luminous members) 
is expected to be low based on the cluster richness \citep{johnston07}.
The richness $N_{200}$ is defined by the number of member galaxies which are 
consistent with the red-sequence color, is within a certain radius of 
the brightest cluster galaxy, and have luminosity above $0.4L_*$ 
\citep[see][for details]{koester07}.  
Although their definition for $N_{200}$ is given only to $z=0.3$, 
we extend their red-sequence color cut to $z=0.35$ based on Table~1 of 
\citet{eisenstein01}, and determine the equivalent richness to be 
$N_{200} = 2$ from their SDSS magnitudes \citep{sdssdr6}.
This is lower than the lowest richness bin available in \citet{johnston07},
an indication that the Q0957 cluster is more of a group than a cluster, 
and hence that the member count is low.

\subsection{CTE Correction}

The galaxy shape is affected by the degradation of the 
charge transfer efficiency (CTE) in the ACS WFC CCD.
When a charge is read out from a CCD pixel, some of the
charge is trapped and released at a later time, causing 
the galaxy image to elongate along the charge read-out direction,
adding an excess shear $\gamma^{\rm cte}_+<0$ along the pixel axis,
where $\gamma_+$ of $>0$ or $<0$ indicates elongation in the horizontal
or vertical direction, respectively (see \S\ref{sec:multipolemoments}
for definition of the orthogonal shears $\gamma_+$ and $\gamma_\times$).
The effect of charge-trailing on the object shape is worse for 
objects that are smaller, have low flux, require larger number of 
transfers to the readout register
(i.e., objects located close to the chip gap in the ACS WFC), and 
exposures that have been taken at a later date.  

We correct for CTE effects using a parametric model given
the flux, size, location on CCD and observation date to estimate the
spurious elongation, and correct for it by removing this shear from 
each object in the shape catalog.
The model was derived from galaxies in the HST/COSMOS Survey
similarly to the one used by \citet{rhodes07}, 
but additionally takes sky background variations into account
\citep[see][for further details]{schrabback09}.

The mean correction averaged over all galaxies in our sample is
$\langle \gamma_+^\mathrm{cte}\rangle = -0.008\pm 0.010$,
with the worst case correction for small, dim objects near the 
chip gap being up to $\gamma_+^\mathrm{cte}=-0.04$.
Given the $2\times2$ mosaic configuration, we expect the 
CTE effect on the measured shear multipole moments
to vanish to first order for the even multipoles.  Overall,
we find the correction to the mass distribution to be negligible, 
of the order $<3\%$ (well within the error bar, see \S\ref{sec:results}) 
for the mass sheet, and within $<10\%$ of 
1$\sigma$ error bar for the multipole moments.

\subsection{Mass Sheet and Multipole Moments}
\label{sec:multipoles}

The two-dimensional multipole mass distribution within an aperture 
can be obtained in a non-parametric fashion from the WL shear information
surrounding the aperture.  The overall mass sheet (i.e., the combined 
average mass sheet from the SL model and the unconstrained cluster term 
$\kappa_c$) corresponds to the monopole.  
Here we summarize the results from \citet{BN08}.

\subsubsection{Mass Sheet}

For any mass distribution,
the expected azimuthally averaged tangential shear at radius $r$ 
from the mass center is \citep{escude91} 
\begin{equation}
  \bar\gamma_t (r) 
= \frac{\bar{\Sigma}(<r)-\bar{\Sigma}(r)}{\Sigma_{\rm crit}}
\equiv \frac{\Delta\Sigma(r)}{\Sigma_{\rm crit}}
= \bar{\kappa}(<r) - \bar{\kappa}(r)
\label{eq:tangentshear}
\end{equation}
where $r$ is the angular distance from the cluster center,
$\bar\gamma_t(r)$ is the tangential shear averaged at $r$,
$\bar{\Sigma}(<r)$ is the mean surface density within an aperture
of radius $r$, $\bar{\Sigma}(r)$ is the azimuthally averaged 
surface density at radius $r$, and $\Sigma_{\rm crit}$ is the critical 
surface density, and $\Delta\Sigma(r)\equiv\Sigma_{\rm crit}\bar\gamma_t$ 
is the value often quoted in WL literature.
The ratio of the surface density to $\Sigma_{\rm crit}$ is the convergence 
$\kappa$.
We can directly estimate the mass sheet within an aperture 
using the aperture mass method \citep{fahlman94}:
the average convergence within a circular aperture of radius $R$ is 
\begin{equation}
 \bar{\kappa}(<R) \equiv \frac{1}{\pi R^2}\int_{r<R} d^2r\,\kappa({\bf r}) 
=\frac{1}{\pi} \int_{r>R} d^2r\,\frac{\gamma_t({\bf r})}{r^2}
\label{eq:masssheet}
\end{equation}
where the tangential shear is integrated over $R<r<\infty$ \citep{KS93}.
The average convergence gives the average surface density
if the critical surface density $\Sigma_{\rm crit}$ is known.  
We discuss $\Sigma_{\rm crit}$ estimation in \S\ref{sec:zestimation}.

The practical estimator for Eq.~(\ref{eq:masssheet}) is a summation
over galaxy shapes
\begin{equation}
  \bar{\kappa}(<R) 
= \frac{1}{\bar{n}\pi{\mathcal R}} \sum_{i \in (r>R)} 
\frac{\gamma_{t,i}}{r^2}
\label{eq:mssummation}
\end{equation}
where $\bar{n}$ is the surface number density of source galaxies and
$\gamma_{t,i}$ is the tangential component of the $i$th galaxy shape.
Here we have assumed that the locally averaged galaxy shape (tangential
component) is an estimator of the shear, 
$\gamma_t \approx \langle\gamma_{t,i}\rangle/{\mathcal R}$,
where ${\mathcal R}$ is the responsivity, the multiplicative
correction factor for the estimation of shear from galaxy shapes \citep{BJ02}.

In practice, the summation over all galaxies at $r>R$ is truncated
at a maximum radius $R_{\rm max}$.  
Equation~(\ref{eq:mssummation}) then becomes
\begin{equation}
  \bar{\kappa}(<R) - \bar{\kappa}(<R_{\rm max})
= \frac{1}{\bar{n}\pi{\mathcal R}} \sum_{i \in (R<r<R_{\rm max})} 
\frac{\gamma_{t,i}}{r^2}
\label{eq:msannulus}
\end{equation}
The truncation at $R_{\rm max}$ is a source of uncertainty; 
that is, only the difference in the mass sheet at different aperture radii
can now be determined (this is the WL mass sheet degeneracy).  
We then assume an appropriate model to derive the mass distribution 
within $R_{\rm max}$, and hence $\bar{\kappa}(<R_{\rm max})$.  
We feel safe in doing so, since this correction is reasonably smaller 
than $\bar{\kappa}(<R)$.

If the WL measurement yields an average convergence of $\bar\kappa$
within $r=R$ (normalized to the quasar redshift) 
and the SL modeling (assuming $\kappa_c=0$ in Eq.~(\ref{eq:galcluster})) 
yields an average convergence of $\bar\kappa_{SL}$,
then the relation of the degenerate mass sheet $\kappa_c$ to the measured
mass sheet $\bar\kappa$ is
$\bar\kappa = \kappa_c + (1-\kappa_c)\bar\kappa_{SL}$, or 
\begin{equation}
  1-\kappa_c = \frac{1-\bar\kappa}{1-\bar\kappa_{SL}}.
\label{eq:kappa}
\end{equation}

\subsubsection{Multipole Moments}
\label{sec:multipolemoments}

The method of aperture mass \citep{fahlman94, schneider96, SB97}, which 
integrates tangential shear to obtain the mass sheet within an aperture, 
can be generalized to obtain the ``interior'' and ``exterior'' multipoles 
defined relative to a circle of radius $R$ \citep{BN08}.
The ``interior monopole'' corresponds to the mass sheet, and is constrained
from the shear information exterior to $r=R$.  The other multipoles are 
constrained in a similar manner, where the shear information exterior to
$r=R$ constrains the mass multipole moments interior to $r=R$, and vice versa.

In order to utilize the shear information to extract mass multipoles, 
first we define the complex shear relative to the tangent to the circle
\citep[see Eq.~(11) of ][]{BN08}
\begin{equation}
  \Gamma({\bf r}) \equiv \gamma_t({\bf r}) + i\gamma_s ({\bf r})
   = -\gamma({\bf r}) e^{-2i\phi}
\label{eq:Gamma}
\end{equation}
where the coordinate origin of ${\bf r}\equiv(r,\phi)$ is 
at the center of the circle, and $\gamma_t$ and $\gamma_s$ are the (real)
tangential and ``skew''\footnote{The conventional name for this term is 
``radial'' shear ($\gamma_r$).  
However, a shear which elongates in the radial direction is simply a 
negative tangential shear, and is not orthogonal to $\gamma_t$.  
The shears orthogonal to the tangential/radial
directions are aligned $\pm$45 degrees from tangential/radial; hence
we flout past conventions and rename this the ``skew'' shear, $\gamma_s$. }
shear components with respect to the polar coordinates, respectively.
$\gamma({\bf r})\equiv \gamma_+({\bf r}) + i\gamma_\times({\bf r})$
is the complex shear defined in the $(x,y)$ coordinates, where
$\gamma_+$ is the shear along the $x$- or $y$-axis, and 
$\gamma_\times$ is the shear along the diagonal halfway between the axes.

The interior and exterior mass multipole moments relative 
to a circle of radius $R$ are defined as
\begin{eqnarray}
Q^{(m)}_{\rm in}(R) & \equiv & \int_{r<R} d^2r\, r^m e^{-im\phi} \kappa({\bf r}) \\
Q^{(m)}_{\rm out}(R) & \equiv & \int_{r>R} d^2r\, r^{-m} e^{im\phi} 
\kappa({\bf r}) 
\end{eqnarray}
where these definitions are normalized to agree with
Eqns~(B5) of \citet{SB97}, but with an alteration in the phase convention.
If the {\em cluster} potential at $r<R$ is described
as Eq.~(\ref{eq:galcluster}), then, assuming 
no contribution from the galaxy potential $\psi_g$,
the interior and exterior multipoles $Q^{(m)}$ are related to 
the cluster constants as
\begin{eqnarray}
Q^{(0)}_{\rm in} (R) & = & \pi R^2 \kappa_c \\
Q^{(1)}_{\rm in} (R) & = & \frac{\pi}{4} R^4 \sigma^*_c\;(1-\kappa_c) \\
Q^{(2)}_{\rm out} (R) & = & - \pi  \gamma_c\; (1-\kappa_c) \\
Q^{(3)}_{\rm out} (R) & = & - \frac{\pi}{2} \delta_c \;(1-\kappa_c)
\end{eqnarray}
independent of the cluster mass distribution at $r>R$, where
$\kappa$ and $\sigma$ are the monopole and dipole mass 
and $\gamma$ and $\delta$ are the constant
and $m=3$ shear terms within $r=R$, respectively, and
$\sigma^*$ indicates complex conjugation. 
The terms $Q^{(2)}_{\rm in}$ and $Q^{(3)}_{\rm in}$ vanish (i.e., 
there are no mass distribution within $R$ of these multipoles), since
$\kappa\equiv\frac{1}{2}\nabla^2\psi=0$ if $\psi\propto r^me^{\pm im\phi}$, 
and the $Q^{(1)}_{\rm out}$ term produces no shear internal to $r<R$.  
These terms have no effect upon the 
lens model or the time delay, and therefore can be ignored.
The $m=-2$ and $m=-3$ shear patterns {\it within} $R$ (corresponding to
$\gamma_c$ and $\delta_c$ terms in Eq.~(\ref{eq:galcluster})) 
are generated by a quadrupole and sextupole mass distribution, respectively, 
external to $R$ \citep[see Fig.~1 of][]{BN08}, hence these terms correspond
to the external mass distributions $Q^{(m)}_{\rm out}$.  
If $\psi_g\neq0$, then the appropriate multipole terms 
$Q^{(m)}_{g,\rm in}$ or $Q^{(m)}_{g,\rm out}$ 
for the galaxy potential enter each of the multipole terms, 
and the WL measurement constrains the sum of the SL galaxy model 
and cluster terms.

The multipoles of the convergence $\kappa$ are related to the 
integrals over the complex shear $\Gamma$, Eq.~(\ref{eq:Gamma}), as 
\begin{eqnarray}
\label{eq:annulus1}
\int_{r>R}d^2r \, \Gamma({\bf r})
r^{-m-2}e^{-im\phi} & = & 
R^{-2m-2} Q^{(m)}_{\rm in}(R) \\
&&\qquad (m\ge 0) \nonumber\\
\label{eq:annulus2}
\int_{r<R}d^2r \, \Gamma({\bf r})
r^{m-2}e^{im\phi} & = & 
 R^{2(m-1)} Q^{(m)}_{\rm out}(R) \\
&& \qquad (m\ge 1) \nonumber 
\end{eqnarray}
where the integral can be converted to a summation over 
source galaxy shapes as an estimator to the integral.  
The practical estimators avoids summation over $R\rightarrow\infty$ 
or in the $r<R$ region (where the weak lensing shear estimator breaks down)
by integrating over an annular region $R_1<r<R_2$:
\begin{eqnarray}
\lefteqn{\frac{1}{\bar{n}{\mathcal R}}\sum_{R_1<r_j<R_2} r_j^{-2}\gamma_{t,j} =}
 \nonumber \\ 
&& \qquad \qquad R^{-2}_1 {Q}^{(0)}_{\rm in}(R_1)
     - R^{-2}_2 {Q}^{(0)}_{\rm in}(R_2)
\label{eq:Q1} \\
\lefteqn{\frac{1}{\bar{n}{\mathcal R}}\sum_{R_1<r_j<R_2} r_j^{-3}(\gamma_{t,j} + i\gamma_{s,j}) e^{-i\phi_j}=} \nonumber \\
&& \qquad \qquad R^{-4}_1 {Q}^{(1)}_{\rm in}(R_1)
     - R^{-4}_2 {Q}^{(1)}_{\rm in}(R_2)
\label{eq:Q2} \\
\lefteqn{\frac{1}{\bar{n}{\mathcal R}}\sum_{R_1<r_j<R_2} (\gamma_{t,j} + i\gamma_{s,j}) \, e^{+2i\phi_j}=} \nonumber \\
&& \qquad \qquad R^2_2 {Q}^{(2)}_{\rm out}(R_2) - R^2_1 {Q}^{(2)}_{\rm out}(R_1) 
\label{eq:Q3} \\
\lefteqn{\frac{1}{\bar{n}{\mathcal R}}\sum_{R_1<r_j<R_2} r_j (\gamma_{t,j} + i\gamma_{s,j}) \,e^{+3i\phi_j} = } \nonumber \\
&& \qquad \qquad R^4_2 {Q}^{(3)}_{\rm out}(R_2) - R^4_1 {Q}^{(3)}_{\rm out}(R_1) 
\label{eq:Q4}
\end{eqnarray}
where the summation is over the $j$th galaxies at radial distance $r_j$
and azimuth angles $\phi_j$.
In Eq.~(\ref{eq:Q1}), the summation over $\gamma_t$ is referred to
as the E-mode aperture mass, which corresponds to the physical quantity
which is the surface mass density, while a summation over $\gamma_s$ 
is the B-mode aperture mass which is expected to vanish, and hence 
provides some measure of systematics present in the monopole 
\citep[similar measures are not available for the higher-order 
multipole terms, as discussed in][]{BN08}.
The multipoles on the right hand sides are that of the combination of 
the cluster and galaxy potential; hence the summation constrains
the {\em combination} of the cluster and galaxy multipoles.

\subsection{Redshift Estimation}
\label{sec:zestimation}

Our mass estimators from tangential shear yields estimates of 
the convergence, but not the mass sheet density $\Sigma$ itself.  
It is important to know the critical surface mass density $\Sigma_{\rm crit}$,
which normalizes the convergence $\kappa \equiv \Sigma/\Sigma_{\rm crit}$ 
and is redshift dependent, since the convergence obtained from WL 
(at a given source redshift distribution) must be converted to an 
appropriate value for SL modeling, where the source is at 
$z_{\rm src}=z_{\rm quasar}=1.41$.

\begin{figure}
\plotone{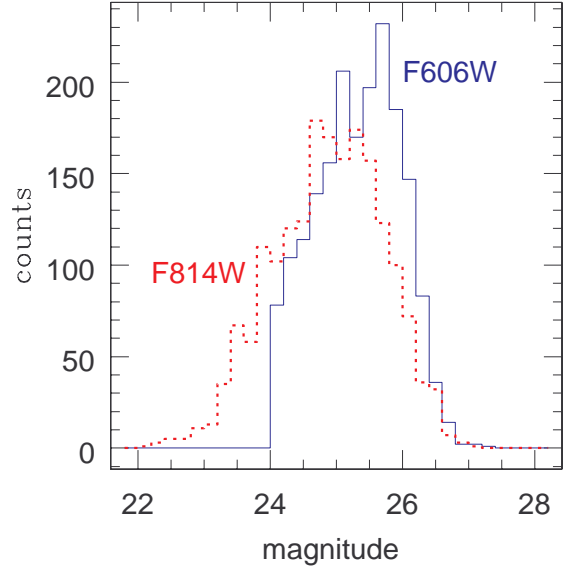}
\caption[Distribution of F606W and F814W AB magnitudes of galaxies used in 
the WL analysis.]
{Distribution of F606W (solid line) and F814W (dotted line)
AB magnitudes of galaxies used in the WL analysis.
The objects were detected in the F606W/F814W combined image, and 
the magnitudes in each were determined using the {\tt SExtractor} 
double-image mode \citep{Bertin}.
The majority (92\%) of our objects have F814W magnitude less than 26.
\label{fig:maghist}}
\end{figure}

In order to obtain the critical surface density $\Sigma_{\rm crit}$
for WL, the source galaxy redshift distribution must be estimated.
We adopt the magnitude dependent parameterization \citep{BF93}
\begin{equation}
\left\{
\begin{array}{l}
\frac{dN}{dz}(z,m) \propto z^2\exp[-(\frac{z}{z_0(m)})^{3/2}]\\
z_0(m) = \frac{z_{\rm med}(m)}{1.412}
\end{array}
\right.
\label{eq:zdist}
\end{equation}
where $\frac{dN}{dz}(z,m)$ is the magnitude dependent redshift distribution, 
and $z_{\rm med}(m)$ is the median redshift as a function of magnitude $m$.
From the COSMOS\footnote{Cosmic Evolution Survey 
({\tt http://cosmos.astro.caltech.edu/}).
COSMOS is an ACS survey over 1.67 deg$^2$, where a single orbit 
($\sim2000$ seconds) exposure is tiled over this relatively wide field
in F814W.  It has 50\% completion for sources $0.5''$ in diameter 
at F814W $I_{\rm AB} =26.0$ \citep{scoville07}.} 
ACS data, \citet{Leauthaud07} obtain the
median redshift as a function of F814W magnitude as 
\begin{equation}
  z_{\rm med} = (0.18\pm 0.01) \times m_{\rm F814W} - (3.3 \pm 0.2)
\label{eq:medianz}
\end{equation}
This relation is valid over the magnitude range $20 < m_{\rm F814W} < 24$;
however, their data suggest that
the relation can be extended out to $m_{\rm F814W} < 26$,
based on the 
UDF\footnote{Hubble Ultra Deep Field ({\tt http://www.stsci.edu/hst/udf/}).
The UDF data is a multi-color, deep image over a single ACS 
field-of-view (11.97 arcmin$^2$) and has a 10-$\sigma$ limiting magnitude 
(for a $0.5''$ diameter aperture) of $28.4$ in the 
F850LP filter, with 144 orbit exposures.
The F850LP filter is narrower and less efficient than the F814W filter.} 
data \citep{Coe06}, which show agreement in the 
$20<m_{\rm F814W}<24$ region with COSMOS data. 
Hence we use Equation~(\ref{eq:medianz}) to estimate our median redshift,
since our objects are within this magnitude range (Figure~\ref{fig:maghist}).

The final redshift distribution was estimated in the following manner:
\begin{enumerate}
\item Divide the galaxies into $\Delta m_{\rm F814W}=0.25$ bins,
\item determine the redshift distribution for each bin
using Eq.~(\ref{eq:zdist}) and $z_{\rm med}$ from Eq.~(\ref{eq:medianz}), and
\item sum over each distribution with each bin properly weighted.
\end{enumerate}
The weight of each bin is determined from the weight each galaxy shape
gets in estimating the shear \citep{BJ02}.
From the estimated source redshift distribution, we calculate the mean 
lensing strength, which is proportional to $\Sigma_{\rm crit}^{-1}$, 
and hence to
\begin{equation}
\left\langle\frac{D_{LS}}{D_{OS}}\right\rangle
 = \frac{\int dm \int dz \, \frac{dN}{dz}(z,m) \, w(m) \frac{D_{LS}}{D_{OS}}}
        {\int dm\,w(m)}
\end{equation}
where $\frac{dN}{dz}(z,m)$ is normalized to unit integral, 
and $w(m)$ is the weight per magnitude bin.
We find $\left\langle\frac{D_{LS}}{D_{OS}}\right\rangle = 0.572$, or
the mean weak lensing critical density to be
\begin{equation}
  \Sigma_{\rm crit} = \frac{c^2}{4\pi G} 
    \frac{1}{D_{OL}}
    \left[\left\langle\frac{D_{LS}}{D_{OS}}\right\rangle\right]^{-1}
   = 3800 h \, M_\sun\,{\rm pc}^{-2}
\end{equation}
for a flat $\Lambda\mathrm{CDM}$ cosmology with $\Omega_\mathrm{m} = 0.24$,
 and where $z_{\rm lens} = 0.355$.

Since the geometry dependence of the gravitational lens system is 
contained in $\Sigma_{\rm crit}$, its value differs for the same lens
if the source object is at different redshift.
We compare the critical density $\Sigma_{\rm crit}$ obtained above to
$\Sigma_{\rm crit}^Q = 3350 h \, M_\sun\,{\rm pc}^{-2}$ for the lensed quasar source at 
$z=1.41$, and find that the lensing signal, which is proportional to
$\Sigma_{\rm crit}^{-1}$, is enhanced by 13\% relative to the WL galaxy population.
In other words,
\begin{equation}
\kappa^Q({\bf r}) 
= \frac{\Sigma_{\rm crit}}{\Sigma_{\rm crit}^Q} \, \kappa({\bf r})
\label{eq:WLtoSL}
\end{equation}
where $\kappa^Q({\bf r})$ is the convergence at quasar redshift, 
$\Sigma_{\rm crit}$ and $\Sigma_{\rm crit}^Q$ are the critical density for sources at
the WL galaxy population and quasar, respectively, and
$\kappa({\bf r})$ is the convergence with respect to the WL galaxy population.

\section{Results}
\label{sec:results}

\subsection{Mass Sheet}

\begin{figure*}
\plotone{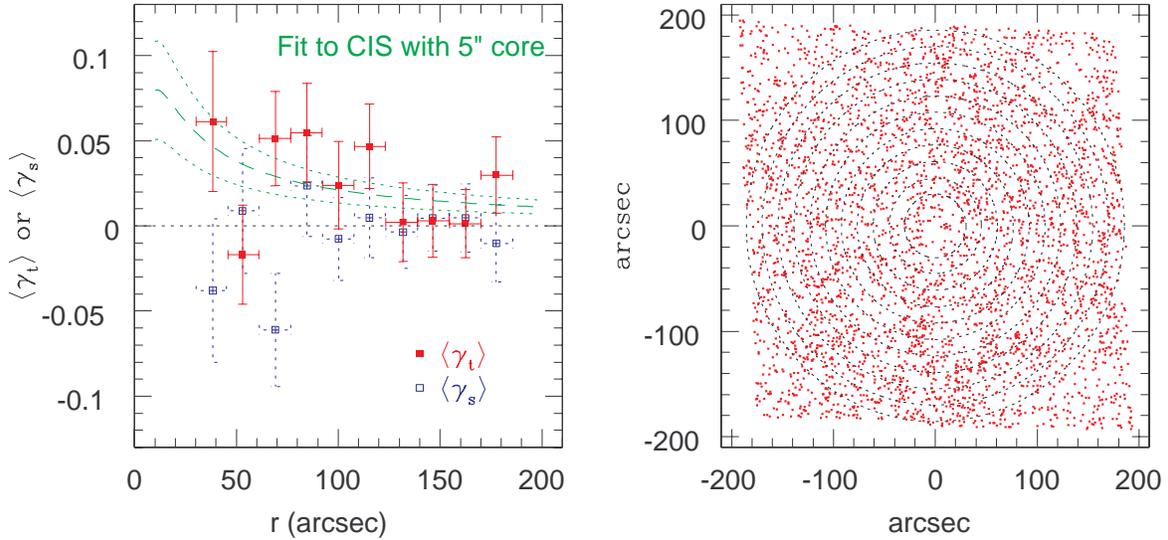}
\caption[Plot of azimuthally averaged tangential shear, 
$\langle \gamma_t \rangle$.]
{{\it Left}: Plot of azimuthally averaged tangential shear, 
$\langle \gamma_t \rangle$, centered 
on G1, as a function of radius $r$ (solid squares).  
The dashed line is the best fit isothermal sphere with $5''$ core, with
the short-dashed lines indicating the one-sigma deviation of the fit.
The B-mode (``skew'') shear $\langle \gamma_s \rangle$ 
(open squares) is consistent with 
zero, as expected from shears generated from a centrally distributed mass.
{\it Right}: Radial bins used in obtaining the mean tangential shear.
Each dot indicates a galaxy in the $6'\times 6'$ ACS mosaic which are 
suitable for use in the weak lensing analysis, 
while the annuli delineated by the dotted line indicate the radial bins.
The radius of the outermost circle is $R_{\rm max}=186''$.
The coordinates are centered around the lensing galaxy G1.
\label{fig:tangentshear}}
\end{figure*}

The average tangential shears at different radii 
are plotted in Figure~\ref{fig:tangentshear}.  
The data points, although noisy, are independent of each other, 
so it is appropriate to fit a model to these data.
We assume a core-softened isothermal sphere (CIS) mass distribution 
\begin{equation}
\kappa({\bf r}) = \frac{\Sigma_0}{\Sigma_{\rm crit}}\;[1+(r/r_c)^2]^{-1/2}
\end{equation}
and use Eq.~(\ref{eq:tangentshear}) to compute the corresponding 
tangential shear.  Fixing the core radius at $r_c=5''$, we find
$\Sigma_0/\Sigma_{\rm crit} = 0.47\pm0.17$, or
$\Sigma_0 = (1800 \pm 600)h\,M_\sun\,{\rm pc}^{-2}$ 
(1 $\sigma$ error estimates), with reduced $\chi^2$ of $0.93$.  
This corresponds to a cluster velocity dispersion of 
$\sigma_v = 420\pm70$ km s$^{-1}$ ($h=0.7$).  
A fit to a NFW model \citep{NFW} was attempted, but did not 
provide any useful constraint on the concentration.

\begin{figure*}
\plotone{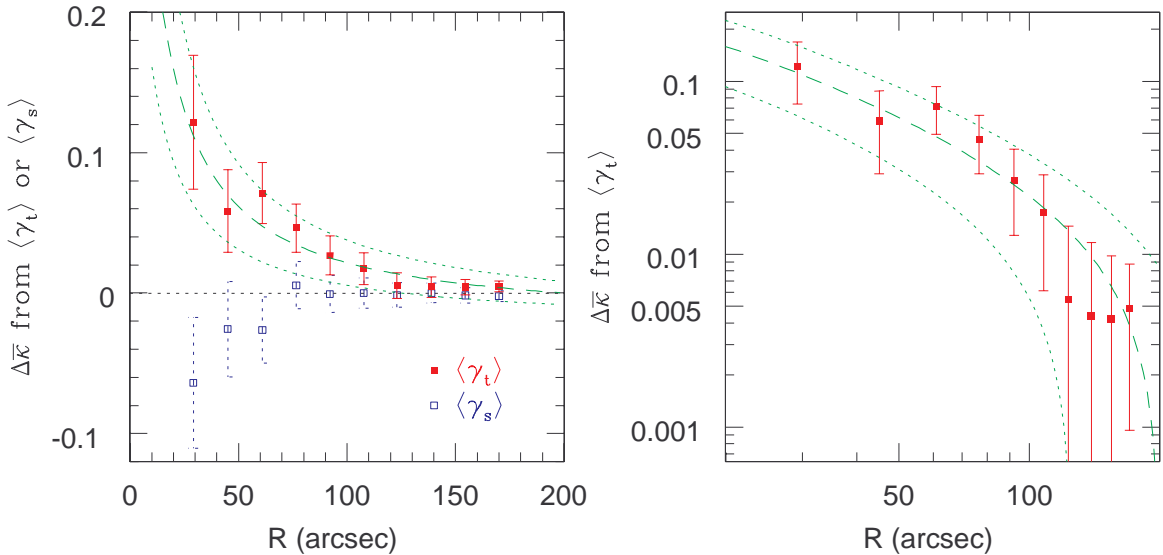}
\caption[The aperture mass statistic of the Q0957+561 cluster.]
{{\it Left}: The aperture mass statistic, which plots the 
differential convergence calculated from integrating over the
tangential shear in Eq.~(\ref{eq:msannulus}).  
The solid squares are the tangential shear integrals, the dashed line
is the fit to an core-softened isothermal sphere with $5''$ core with 
the dotted lines indicating
one-sigma deviations of the fit, and the open squares 
are the imaginary components (B-mode aperture masses, or 
azimuthally averaged skew shears)
whose signal should be consistent with zero.
The points in this figure are correlated,
since each point is an integral over the annulus from $R_{\rm max}$ 
to the inner radius $R$.
{\it Right}: The same plot in log scale.
\label{fig:map}}
\end{figure*}

Figure~\ref{fig:map} shows the aperture mass statistic as defined in 
Eq.~(\ref{eq:msannulus}), with respect to a varying inner annulus. 
The points in this figure are correlated, since each point is
an integral over galaxies in the annulus from $R_{\rm max}=186''$ 
to the inner radii $R$.  The points show the azimuthally averaged
radial profile of the mass concentration around Q0957.  Because of the
WL mass sheet degeneracy, it is not the true radial profile, but
a relative value with respect to a mass sheet averaged within a 
circular aperture of $r<R_{\rm max}$.

From the aperture mass, we find the average convergence overdensity to be
$$
\Delta\bar\kappa=\bar\kappa\,(<30\arcsec) - \bar\kappa\,(<186\arcsec)
 = 0.122\pm0.048  
$$
where the radii $R=30\arcsec$ and $R_{\rm max}=186\arcsec$ are 
with respect to G1.
The average convergence within $R_{\rm max}=186\arcsec$ can be estimated
from the CIS model, whose fit value yields
$\bar\kappa\,(<186\arcsec) = 0.024\pm0.012$.  The error here is 
conservative (i.e., not based on the error in the fit) to account for
various possible mass distributions.
The two results can then be combined to yield
$$
  \bar\kappa\,(<30\arcsec) = 0.146\pm0.049
$$
From the redshift distribution of the WL source galaxies,
this convergence corresponds to a mean mass sheet density of
$$
  \bar\Sigma\,(<30\arcsec) = (550\pm190)h \,M_\sun{\rm pc}^{-2}
$$
For sources at the quasar redshift $z=1.41$,
this mass sheet then corresponds to a convergence of 
$$
\bar\kappa(<30\arcsec) = 
0.166\pm0.056
$$
for the strong lensing analysis (see Eq~(\ref{eq:WLtoSL})).
As explicitly stated above, this ``mean mass sheet'' includes 
the G1 mass averaged over the $r<30''$ disk; 
hence to obtain the cluster mass sheet $\kappa_c$, the galaxy mass 
(modeled from SL) must be properly removed from the above quantity
\citep[Eq.~(\ref{eq:kappa});][]{fadely09}.

\subsubsection{Comparison with Fischer \etal}

Our result is consistent with the WL results based on
Canada France Hawaii Telescope (CFHT) images from \citet{FischerProFit}.
They obtain $\bar\kappa(<30\arcsec)=0.16\pm 0.05$ (from their Figure~6) 
compared to our $\bar\kappa(<30\arcsec)=0.146\pm 0.049$, where we have 
compared the shears as obtained from the galaxy shapes, without normalizing
to the quasar redshift.  
While the two results and error magnitudes are nearly identical, 
they are obtained in a completely different fashion.  
We obtain the shear calibration factor (our responsivity ${\mathcal R}$) 
and propagate the shape noise and shape measurement errors from individual 
galaxies into the estimated shear in a deterministic manner based on 
the formalisms of \citet{BJ02}, while Fischer \etal\ calibrate 
the shear and obtain the error based on Monte-Carlo simulations.  

The similarity in the error magnitude could possibly be 
attributed to the similar number of galaxies we use (1651 and 1866 galaxies
for Fischer \etal\ and this study, respectively), since shear estimate error 
is typically dominated by the statistical shape noise.  Both studies
use source galaxies in a similar area and magnitude range.
However, the resolution of the images used in the two studies are 
vastly different: 
the CFHT images have $0.6\arcsec$ FWHM seeing with 0.207\arcsec/pixel, 
compared to our $0.13\arcsec$ FWHM with with 0.05\arcsec/pixel.  
Since the images with the smaller seeing and pixel scale should reveal
the galaxy shapes better, the similarity in the results indicates that
the error is indeed dominated by the statistical error.
Therefore, in order to improve upon the mass sheet precision, we would need to 
increase the number count of the source galaxies, which would be possible
if source galaxies of lower S/N in the HST images can be utilized.
Currently, the accuracy in our galaxy shape measurement and shear 
estimation method is limited to high S/N objects ($>25$). 
Hence a shape measurement method which can be shown to be 
accurate at low S/N would improve precision in the mass sheet measurement, 
given the same HST images.  

Although the two observed convergence are consistent with each other,
\citet{BF99} report a convergence of $\bar\kappa(<30\arcsec)=0.26\pm 0.08$ 
based on the results of \citet{FischerProFit}, 
when normalized to the quasar redshift. 
This discrepancy comes from the difference in the assumed 
$\Sigma_{\rm crit}$ values: although the redshift of the quasar
has not changed, both the assumed WL source galaxy distribution 
and the cosmology are different between the two analyses.  
While we use a magnitude-based empirical redshift distribution
(\S\ref{sec:zestimation}) with a flat $\Lambda$CDM with $\Omega_m=0.24$,
\citet{FischerProFit} estimate the redshift distribution based on 
Monte-Carlo simulation of an assumed redshift evolution of galaxy size,
along with an open CDM cosmology with $\Omega_m=0.1$.

\subsubsection{Comparison with Other Cluster Mass Estimates}

\citet{chartas02} use $2'\times2'$ Chandra observation to 
determine the mass distribution of the cluster.  
From the spatial distribution of the X-ray luminosity and the 
X-ray temperature, they find the cluster mass to be 
$5.0^{+1.3}_{-2.0}\times10^{13} M_\sun$ within a radius of 
0.5 $h_{75}^{-1}$ Mpc ($\sim2'$) of the cluster center.
Our corresponding WL mass estimate within the same radius is 
$(3.3\pm1.1)\times10^{13} M_\sun$ (for $h=0.7$),
an agreement within $\sim1\sigma$.  We note that since
the mass here is an integral over a radial density profile,
the cluster mass estimate agreement between WL and X-ray, which 
assumes a different radial profile (CIS and $\beta$-model, 
respectively), will differ depending on the outermost radius chosen.  

On the other hand, \citet{garrett92} and \citet{angonin94} obtain
a cluster member velocity dispersion of $\sigma_v=715\pm130$ km s$^{-1}$
from redshifts for 21 probable cluster members, compared to our
$\sigma_v$ equivalent of $420\pm70$km s$^{-1}$ ($h=0.7$). 
The dispersion data correspond to cluster mass of 
$(17.8\pm6.5)\times10^{13}M_\sun$ ($h=0.7$) within a 
$0.5 h_{75}^{-1}$ Mpc radius.
The $2\sigma$ discrepancy between the WL and velocity-dispersion mass
estimates suggests that the velocity dispersion is possibly highly 
anisotropic, with an enhanced peculiar velocity along the
line of sight.  
\citet{angonin94} suggest that cluster member selection could also be
the cause for the large dispersion, where they find removing a single galaxy at
the edge of the velocity distribution reduces the velocity dispersion to
660 km/s.

\subsection{Multipole Moments}

\begin{figure*}
\plotone{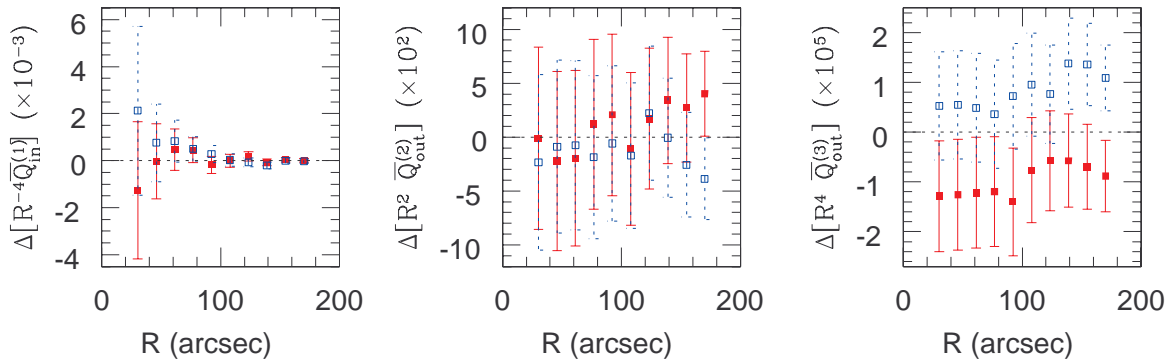}
\caption[Differential multipole moments of Q0957+561 cluster, centered at 
G1.]
{{\it Left}: Differential $R^{-4} {Q}^{(1)}_{\rm in}(R)$ with respect to
the value at $R_{\rm max}$. 
{\it Center}: Differential $R^2 {Q}^{(2)}_{\rm out}(R)$.
{\it Right}: Differential $R^4 {Q}^{(3)}_{\rm out}(R)$.
In all three figures, the solid squares indicate the real components, 
while the open squares indicate the imaginary components.  
The complex numbers encode the multipole orientation, where the 
axes are aligned with West (``$x$'') and North (``$y$'').
The internal mass dipole term has a directionality that points 
approximately to the Northeast with respect to G1, consistent with previous 
studies \citep{FischerProFit,chartas02}.
\label{fig:multipole}}
\end{figure*}

The rest of the (differential) multipole moments as obtained from 
Eqns.~(\ref{eq:Q2}--\ref{eq:Q4}) 
are listed in Table~\ref{tab:multipole}, where the
differential is between the values at $R=30''$ and $R_{\rm max}=186''$.
Figure~\ref{fig:multipole} plots the multipoles
with respect to the value at $R_{\rm max}$ for various inner radii $R$,
which highlights the problem with constraining the $Q^{(m)}_{\rm out}$ terms.
Due to the powers of $R$ involved in the summation, 
the shear signal, as well as the shape noise, are weighed by $R^{-2m-2}$  
for $Q^{(m)}_{\rm in}$,  and by $R^{2m-2}$ for $Q^{(m)}_{\rm out}$. 
For $Q^{(m)}_{\rm in}$, the outer region (dominated by shape noise) are 
weighted less, and the signal at smaller radii can be seen 
(Fig.~\ref{fig:multipole} left).  
At $Q^{(m)}_{\rm out}$, $m>1$, (Fig.~\ref{fig:multipole} center and right),
whatever signal 
existing at the inner radii is overwhelmed by the shape noise in the outer 
radii, and hence the differential signal as $R$ decreases to $30''$ is 
merely noise which dominates at larger radii.  This effect is more severe for
$Q^{(3)}_{\rm out}$ (Fig.~\ref{fig:multipole} right), 
where higher powers of $R$ are involved,
that the shear from $R<100''$ hardly contributes to the integrated shear.

If we are to constrain the multipoles at $R=30''$,
it is also necessary to have a handle on their values at $R_{\rm max}$.
This value is estimated to be up to $20\%$ of the values at $R=30''$
for the $Q^{(2)}_{\rm out}$ and $Q^{(3)}_{\rm out}$  terms, and 
around $1\%$ for the $Q^{(1)}_{\rm in}$ term, for a CIS cluster 
whose center is displaced from the G1 center by $5''$ to $30''$.  
Since our multipole signals are already significantly smaller
than the measurement errors, we deduce that the correction from the 
$R_{\rm max}$ term is significantly smaller than the errors
if it arises from a cluster that resembles a CIS profile beyond $R_{\rm max}$.

\begin{table*}
\center
\begin{tabular}{rcll}
\hline\hline
$R^{-4}_1 Q^{(1)}_{\rm in}(R_1) - R^{-4}_2 Q^{(1)}_{\rm in}(R_2)$
& = & $[(-1.2\pm2.9) + i(+2.1\pm3.6)]\times 10^{-3}$ & arcsec$^{-1}$\\
$R^2_2 Q^{(2)}_{\rm out}(R_2) - R^2_1 Q^{(2)}_{\rm out}(R_1)$
& = & $[(+0.1\pm8.5) + i (-2.3\pm8.2)]\times 10^{2}$ & arcsec$^{2}$ (*)\\ 
$R^4_2 Q^{(3)}_{\rm out}(R_2) - R^4_1 Q^{(3)}_{\rm out}(R_1)$
& = & $[(-1.3\pm1.1) + i (-0.5\pm1.1)]\times 10^{5}$ & arcsec$^{3}$\\ 
\hline\hline
\end{tabular}
\caption[Table of multipole values in cluster Q0957+561.]
{Table of multipole values, where $R_2 = R_{\rm max} = 186''$, and
$R_1 = 30''$.
These values correspond to the innermost points 
plotted in Figure~\ref{fig:multipole}, and hence are normalized to the
WL source redshifts.  
(*) There is an alternative means of obtaining the 
$\gamma_c$ [$R^2Q^{(2)}_{\rm out}$] term, via straightforward averaging of 
the weak shear values within the annulus around $R=R_1$ (see text).
\label{tab:multipole}}
\end{table*}

Ignoring the $R_{\rm max}$ terms from the cluster and 
assuming the lensing potential as stated in Eq.~(\ref{eq:galcluster}),
the cluster expansion coefficients (normalized to the quasar redshift) 
are constrained to
\begin{eqnarray*}
\lefteqn{(1-\kappa_c)\,\sigma_c
 +  \frac{4}{\pi} (1-\kappa_c) 
\left[R^{-4}_1 Q^{(1)}_{{\rm in},g}(R_1) 
 - R^{-4}_2 Q^{(1)}_{{\rm in},g}(R_2)\right]} 
\nonumber \\
&& \ \ = (-0.0018\pm0.0042) + i (-0.0031\pm0.0052) \ \ \ {\rm arcsec}^{-1}
\\
\lefteqn{(1-\kappa_c)\,\gamma_c
 +  \frac{1}{\pi} (1-\kappa_c) 
\left[ Q^{(2)}_{{\rm out},g}(R_1) 
 - \left(\frac{R_2}{R_1}\right)^2 Q^{(2)}_{{\rm out},g}(R_2)\right] }
\nonumber\\
&& \ \  = (0.00\pm0.34) + i (+0.09\pm0.33) \\ 
\lefteqn{(1-\kappa_c)\,\delta_c
 +  \frac{2}{\pi} (1-\kappa_c) 
\left[Q^{(3)}_{{\rm out},g}(R_1) 
 - \left(\frac{R_2}{R_1}\right)^4 Q^{(3)}_{{\rm out},g}(R_2)\right] }
\nonumber\\
&& \ \ =  (+0.115\pm0.098) + i (-0.047\pm0.097) \ \ \ {\rm arcsec}^{-1}
\end{eqnarray*}
As explicitly stated above, 
the galaxy model contribution $Q_{\rm g}^{(m)}$ 
to each moment must be subtracted from the results 
to obtain the cluster moments $\gamma_c, \sigma_c$, and $\delta_c$; 
the measurements constrain only the sum of the galaxy and cluster 
contributions.

As seen in Figure~\ref{fig:multipole}, the error in the $\gamma_c$ and 
$\delta_c$ terms do not converge to any sufficient degree.
However, we attempt to constrain $\gamma_c$ by the following means:
Since this term is simply the constant shear across $r<30''$,
we measure the average of the (WL) shear within this region.
However, since the central portion of this region is strongly lensed,
we average the shear signal over an annular region of $20''<r<40''$ 
to estimate the constant shear within $r<30''$.  
We find the average shear, normalized to the quasar redshift, to be
\begin{eqnarray*}
(1-\kappa_c)\,\gamma_c
& + & \frac{1}{\pi} (1-\kappa_c) 
\left[ Q^{(2)}_{{\rm out},g}(R_1) 
 - \left(\frac{R_2}{R_1}\right)^2 Q^{(2)}_{{\rm out},g}(R_2)\right] \\
& = & (-0.009 \pm 0.045) + i(+0.092 \pm 0.045).
\end{eqnarray*}
This allows us to shrink the error bar considerably, and by doing so
we have assumed that $\gamma_c$ is generated exclusively by mass
exterior to $40''$.  The directionality of the constant shear term
is consistent with an exterior mass quadrupole distribution with 
positive weight along the Northeast/Southwest direction with respect to G1.

\section{Conclusion}

The Q0957+561 gravitational lens system allows for a 
determination of the Hubble constant $H_0$, based on 
its firm time delay between the double images of the quasar; 
the remaining uncertainty originates from our knowledge of the 
lens mass distribution.
The lens consists of the galaxy G1 and the 0957+561 cluster to
which the galaxy belongs.
WL offers complementary constraints to the lens mass 
provided by the SL modeling analysis, by providing the mass sheet and 
multipole moments of the underlying cluster.
We have developed and utilized the formalism for an exact solution for 
aperture mass multipoles from WL shear data in the thin-lens approximation.

The mean convergence in the SL region $r<30''$ is estimated to be
$\bar\kappa=0.166\pm0.056$ (1 $\sigma$) normalized to the quasar redshift,
based on shear data from the annular region $30''<r<186''$,
where the radii are centered on the lens galaxy G1.  
Although our shear measurements are consistent with \citet{FischerProFit}, 
the new quasar-normalized $\bar\kappa$ is more reliable since they are 
based on better data, a more accurate galaxy redshift distribution, and 
a standard cosmology.  
Our results give a $7\%$ precision in the mass sheet degeneracy term 
$(1-\overline{\kappa})$.

The uncertainty in the external multipole terms $\gamma_c$ and $\delta_c$
are too large to provide useful constraints to 
the lens potential $\psi$ within $r<30''$.
However, the constant shear term $(1-\bar\kappa)\gamma_c$ within $r<30''$
can be estimated by straightforward averaging of the shear in the WL region
at $20''<r<40''$.
The internal dipole $(1-\bar\kappa)|\sigma_c|$ 
(whose galaxy contribution is expected to be negligible) 
has a value of $0.006\pm0.006$, compared to 
the Monte Carlo parametric fit values $\sim0.011\pm0.007$ 
obtained by \citet{keeton00}.
The implications for $H_0$ with the full SL lens modeling
are discussed in an accompanying paper \citep{fadely09}.

\acknowledgements {
This work has been supported by grant HST-GO-10569 from the Space Telescope
Science Institute, which is operated by the Association of Universities
for Research in Astronomy, Inc., under NASA contract NAS5-26555.
RN and GMB acknowledge additional support from NASA grant BEFS-04-0014-0018.
TS acknowledges financial support from the Netherlands Organization for 
Scientific Research (NWO) and the German Ministry for Education and Research
(BMBF) through TR33.
}


\end{document}